# Contact Angle Adjustment in Equation of States Based Pseudo-Potential Model

Anjie Hu [a,b]∗, Longjian Li [a], Rizwan Uddin [c]

a. Key Laboratory of Low-grade Energy Utilization Technologies and Systems of Ministry of Education, Chongqing University, Chongqing 400030, PR China

b. School of Civil Engineering and Architecture, Southwest University of Science and Technology, Mianyang 621010, PR China

c. Department of Nuclear, Plasma and Radiological Engineering, University of Illinois at Urbana Champaign, Urbana, IL 61801, USA

**Abstract**

Single component pseudo-potential lattice Boltzmann model has been widely applied in multiphase simulation due to its simplicity and stability. In many research, it has been claimed that this model can be stable for density ratios larger than 1000. However, the application of the model is still limited to small density ratios when the contact angle is considered. The reason is that the original contact angle adjustment method influences the stability of the model. Moreover, simulation results in present work show that, by applying the original contact angle adjustment method, the density distribution near the wall is artificially changed, and the contact angle is dependent on the surface tension. Hence, it is very inconvenient to apply this method with a fixed contact angle, and the accuracy of the model cannot be guaranteed. To solve these problems, a contact angle adjustment method based on the geometry analysis is proposed and numerically compared with the original method. Simulation results show that, with the new contact angle adjustment method, the stability of the model is highly improved when the density ratio is relatively large, and it is independent on the surface tension.

---

∗ Corresponding author. Tel.: +86 15210623987
  E-mail address: anjie@cqu.edu.cn (A. Hu)

## 1 Introduction

Droplets movement on a solid wall is a common and important phenomenon in nature. It plays an important role in many engineering applications such as crude oil attached to rocks, bubbles detachment on the wall in a boiling system, waterproof materials and so on. The key factor of this phenomenon is the wettability of liquid which is directly reflected by the contact angle of the stationary droplet on the wall. This crucial phenomenon can be studied by different numerical methods.

One of these methods is lattice Boltzmann equation (LBE) method [1], also known as lattice Boltzmann method (LBM). Due to its mesoscopic back ground, simplicity and strict second-order accuracy, it has attracted much attention in recent years. Soon after its appearance, several multiphase LBE models have been developed. These models can be summarized into four categories [2]: color models [3], pseudo-potential models [4-6], free energy models [7, 8] and kinetic models [9-12]. Among them, the pseudo-potential models [4-6, 13, 14] have been widely applied because of their simplicity and potential ability to simulate multi-phase problems with large density ratio. Early pseudo-potential models suffer some drawbacks such as numerical instability, spurious velocities and untunable surface tension. However, most of these drawbacks have been overcome or reduced in recent research [15-17]. By properly controlling the mechanical stability condition and adjusting the scale of the equations of state (EOS), the pseudo-potential models now can be applied in large density ratio simulation with relatively high accuracy force methods [16, 17]. Surface tension adjustment can also be achieved by modifying the pressure tensors of these models [15].

Pseudo-potential models have been applied in the contact line problems since it was proposed. Martys and Chen [18] first proposed the method to simulate contact line phenomenon by introducing an interaction force between the fluid and the wall. This method has been further applied in many applications [19] of pseudo-potential simulations: Fan et al. [20] studied the function between the apparent contact angle and the velocity of displacement in a channel with the multi-component pseudo-potential model; Kang et al. [21] further studied the droplet movement in a channel under the influence of the wettability of the fluid; later Benzi et al. [22] applied the method in bubble growth simulation with the single component multiphase pseudo-potential model. Now, simulating contact angle by introducing interaction force between the fluid and the wall has become the main strategy to apply the pseudo-potential model in contact line phenomenon simulations.

However, this method suffers from some limitations mainly concerning contact angle prescription and its influence on the stability of the model. Since it is difficult to quantize the surface tension between fluid and wall, the clear expression of contact angle cannot be obtained. Sukop and Thorne [19] have analyzed the relation between the parameters and the contact angle and derived an analytical expression of the relation for single component pseudo-potential model, but it can only get an approximate prediction when the contact angles are equal to 0 ,90 and 180 degrees, other values of the angle can only be obtained by numerical tests. Huang et al. [23] also proposed an expression to prescribe the contact angle for multi-component pseudo-potential model, although it is clearer than previous methods, it is still an approximate method. Another problem of this contact angle method is that it influences the stability of the model. One of the advantages of single component pseudo-potential model is its ability to simulate large density ratio problems [14]. However, the study in present work found that the interaction force between the fluid and the wall may influence the stability of the model and reduce the range of contact angle adjustment when the density ratio is relatively large.

Recently, Ding et al. [24] proposed a geometrical method to get the prescribed contact angle in the phase-field models. Later, Huang et al. [25] incorporated this method in the He-Chen-Zhang model [10] and Lee-Lin [26] model to overcome the problem that contact angles resulted from surface-energy approach do not precisely agree with the prescribed ones.

To overcome these drawbacks of pseudo-potential model in contact angle simulations, in present paper, we here improve geometrical formulation proposed by Ding et al. [24] and apply it in the pseudo-potential model. The improved contact angle adjustment method is then numerically compared with the original method proposed by Martys and Chen [18]. In order to maintain the stability and investigate the influence of the surface tension on the method, the Kupershtokh's interparticle interaction force format is adapted here, and the pressure tensor modifying surface tension adjustment method [15, 30] is adopted in the MRT operator.

The rest of the paper is organized as follows. Section 2 describes the mathematical theory of the present pseudo-potential model. The effects of the previous contact angle method on the stability will be discussed in Sec. 3. The new method is proposed and numerically compared in Sec. 4. Finally, a brief conclusion will be made in Sec. 5.

**2 Pseudo-potential model**

In the LBE method, the motion of the fluid is described by evolution of the density distribution function. The evolution equation can be written in the form of MRT operator [2, 27] as

$$f_\alpha(\mathbf{x}+\mathbf{e}_\alpha \Delta t, t+\Delta t) - f_\alpha(\mathbf{x},t) = -\left(\mathbf{M}^{-1}\mathbf{\Lambda}\mathbf{M}\right)_{\alpha\beta}(f_\alpha(\mathbf{x},t) - f_\alpha^{eq}(\mathbf{x},t)) + \delta_t \mathbf{F}_\alpha, \quad (1)$$

where $f_\alpha(\mathbf{x},t)$ is the mass distribution function of particles at node $\mathbf{x}$, time $t$; $\mathbf{e}_\alpha$ is the velocity where $\alpha = 0, 1, 2 \cdots N$; $f_\alpha^{eq}(\mathbf{x},t)$ is the equilibrium distribution. The right side of the equation is a collision operator, and $\mathbf{M}^{-1}\mathbf{\Lambda}\mathbf{M}$ is the collision matrix, in which $\mathbf{M}$ is the orthogonal transformation matrix and $\mathbf{\Lambda}$ is a diagonal matrix which is given by (D2Q9)

$$\mathbf{\Lambda} = diag(\tau_\rho^{-1}, \tau_e^{-1}, \tau_\xi^{-1}, \tau_j^{-1}, \tau_q^{-1}, \tau_j^{-1}, \tau_q^{-1}, \tau_\upsilon^{-1}, \tau_\upsilon^{-1},). \quad (2)$$

where $\tau$ represents the relax time, $\tau_\upsilon^{-1}$ is related to the viscosity of the fluid $\upsilon$, the relationship between the relax time and the viscosity can be written as

$\upsilon = \frac{1}{3}\left(\tau_\upsilon - \frac{1}{2}\right)\Delta t$. $\mathbf{F}_\alpha$ is the force term which is given by

$$\mathbf{F}_\alpha = \mathbf{M}^{-1}\left(\mathbf{I} - \frac{1}{2}\mathbf{\Lambda}\right)\mathbf{M}\overline{\mathbf{F}}. \tag{3}$$

For MRT operator, the collision is calculated in the moment space. The density distribution function and its equilibrium distribution function can be transferred into moment space by $\mathbf{m} = \mathbf{Mf}$ and $\mathbf{m}^{eq} = \mathbf{Mf}^{eq}$. For the D2Q9 lattice, the equilibria $\mathbf{m}^{eq}$ is given by

$$\mathbf{m}^{eq} = \rho\left(1, -2 + 3|v|^2, 1 - 3|v|^2, v_x, -v_x, v_y, -v_y, v_x^2 - v_y^2, v_x v_y\right). \tag{4}$$

The force terms in moment space can be written as

$$\tilde{\mathbf{F}} = \mathbf{MF} = \left(\mathbf{I} - \frac{1}{2}\mathbf{\Lambda}\right)\mathbf{M}\overline{\mathbf{F}} = \left(\mathbf{I} - \frac{1}{2}\mathbf{\Lambda}\right)\mathbf{S}, \tag{5}$$

where $\mathbf{S}$ is given by

$$\mathbf{S} = \begin{bmatrix} 0 \\ 6(v_x F_x + v_y F_y) \\ -6(v_x F_x + v_y F_y) \\ F_x \\ -F_x \\ F_y \\ -F_y \\ 2(v_x F_x - v_y F_y) \\ (v_x F_x + v_y F_y) \end{bmatrix}. \tag{6}$$

In the pseudo-potential model, the total force is generally given by the summation of three forces:

$$\mathbf{F} = \mathbf{F}_i + \mathbf{F}_g + \mathbf{F}_s, \tag{7}$$

where $\mathbf{F}_i$ is the interaction force of the fluid, $\mathbf{F}_g$ is the body force and $\mathbf{F}_s$ is the interaction force between the fluid and the solid wall.

For interaction between the nearest neighbors, the interaction force can be generally calculated in two formats. The first is the effective density type proposed by Shan and Chen which can be written as [6]

$$F_{i1}(x,t) = -G\psi(x,t)\sum_{\alpha} w(|e_\alpha|^2)\psi(x+e_\alpha\delta_t,t), \quad (7)$$

where $G$ is the interaction strength, $w(|e_\alpha|^2)$ are the weights, and $\psi(x,t)$ is the effective density. The weights $w(|e_\alpha|^2)$ are $w(1)=1/3$ and $w(2)=1/12$. The second one is potential function type proposed by Zhang et al. [28], which can be written as

$$F_{i2}(x,t) = -\sum_{\alpha} w(|e_\alpha|^2)U(x+e_\alpha\delta_t,t), \quad (8)$$

where $U(x,t)$ is the potential function which is equal to $G\psi^2(x,t)/2$.

To improve the stability of the pseudo-potential model, Kupershtokh et al. [13] proposed a hybrid model by combining these two models mentioned above, which is given by

$$F_{i3}(x,t) = -A\sum_{\alpha=0}^{8}\omega_\alpha U(x+e_\alpha)e_\alpha - (1-A)G\psi(x,t)\sum_{\alpha=0}^{8}\omega_\alpha\psi(x+e_\alpha,t)e_\alpha. \quad (9)$$

In practice, both of effective density and potential function models can be obtained by introducing a non-ideal EOS:

$$p_0 = c_s^2\rho + cG[\psi(\rho)]^2/2 = c_s^2\rho + cU(\rho). \quad (10)$$

To easily control the density distribution, the self-tuning EOS proposed by Colosqui et al. [29] is adopted in present work. The EOS is given by

$$p(\rho) = \begin{cases} \rho\theta_v & \text{if } \rho \leq \rho_1 \\ \rho_1\theta_v + (\rho-\rho_1)\theta_m & \text{if } \rho_1 \leq \rho \leq \rho_2, \\ \rho_1\theta_v + (\rho_2-\rho_1)\theta_m + (\rho-\rho_2)\theta_l & \text{if } \rho > \rho_2 \end{cases} \quad (11)$$

where $\sqrt{\theta_v} = \sqrt{(\partial p/\partial \rho)_v}$ and $\sqrt{\theta_l} = \sqrt{(\partial p/\partial \rho)_l}$ are the speeds of sound of the vapor-phase and liquid-phase, respectively. And $\theta_m$ is the slope in the unstable branch ($\partial p/\partial \rho < 0$). The unknown variables $\rho_1$ and $\rho_2$ are obtained by solving a set of two equations: one for mechanical equilibrium:

$$\int_{\rho_v}^{\rho_l} dp = (\rho_1 - \rho_v)\theta_v + (\rho_2 - \rho_1)\theta_m + (\rho_l - \rho_2)\theta_l = 0, \tag{12}$$

and the other for chemical equilibrium:

$$\int_{\rho_v}^{\rho_l} \frac{1}{\rho} dp = \log(\rho_1/\rho_v)\theta_v + \log(\rho_2/\rho_1)\theta_m + \log(\rho_l/\rho_2)\theta_l = 0, \tag{13}$$

where $\rho_v$ and $\rho_l$ represent the density of vapor and liquid, respectively.

To simulate the vapor-liquid two-phase flow with the influence of a solid surface, Martys and Chen [18] introduced an interaction force between the fluid and the solid wall. This force is generally given by

$$\mathbf{F}_s(\mathbf{x},t) = -G_w \psi(x,t) \sum_\alpha w(|\mathbf{e}_\alpha|^2) S(\mathbf{x} + \mathbf{e}_\alpha \delta_t, t) \mathbf{e}_\alpha \delta_t, \tag{14}$$

where $S(\mathbf{x} + \mathbf{e}_\alpha \delta_t, t)$ is a binary function (it is equal to 1 for solid and 0 for fluid nodes). The parameter $G_w$ controls the strength of the intermolecular force between wall and fluid, therefore it can influence the wettability of the wall. The relationship between the contact angle and the parameter $G_w$ is always obtained by numerical simulation tests with different $G_w$.

To adjust the surface tension, we adopted the pressure tensor modification method in the model. Additional terms were introduced in the Navies-Stokes equation by modifying the MRT LB equation:

$$\mathbf{m}^* = \mathbf{m} - \boldsymbol{\Lambda}(\mathbf{m} - \mathbf{m}^{eq}) + \delta_t \left(\mathbf{I} - \frac{\boldsymbol{\Lambda}}{2}\right)\mathbf{S} + \delta_t \mathbf{C}, \tag{15}$$

where the source term **C** is given by [15, 30]

$$\mathbf{C} = \begin{bmatrix} 0 \\ 1.5\tau_e^{-1}(Q_{xx}+Q_{yy}) \\ Q_3 \\ 0 \\ 0 \\ 0 \\ 0 \\ -1.5\tau^{-1}(Q_{xx}-Q_{yy}) \\ -\tau_v^{-1}Q_{xy} \end{bmatrix}, \qquad (16)$$

where $Q_3$ does not influence the simulation results, here we choose $Q_3 = 0$. The modification tensor is given by [30]

$$\mathbf{Q} = \kappa\left[(1-A)\psi(\mathbf{x})\sum_{\alpha=1}^{N}w(|\mathbf{e}_\alpha|^2)(\psi(\mathbf{x}+\mathbf{e}_\alpha)-\psi(\mathbf{x}))\mathbf{e}_\alpha\mathbf{e}_\alpha + \frac{1}{2}A\sum_{\alpha=1}^{N}w(|\mathbf{e}_\alpha|^2)(\psi(\mathbf{x}+\mathbf{e}_\alpha)^2-\psi(\mathbf{x})^2)\mathbf{e}_\alpha\mathbf{e}_\alpha\right]$$

. (17)

The surface tension in the model is proportional to $1-\kappa$ when the viscosities of both phases are even.

The contact angle is given by the Young's equation:

$$\cos\theta_1 = \frac{\sigma_{sl}-\sigma_{sv}}{\sigma_{lv}}, \qquad (18)$$

where $\sigma_{lv}$ is the surface tension between vapor and liquid phases, and $\sigma_{sv}$, $\sigma_{sl}$ are the surface tension between the wall and the vapor phase and the surface tension between the wall and the liquid phase, respectively. For pseudo-potential model, it is difficult to calculate these surface tensions, so the contact angle is always obtained by numerical tests.

## 3 Contact angle simulation with Martys and Chen's solid-fluid interaction force

In this section, the contact angle adjustment method proposed by Martys and Chen (Eq. (18)) is numerically analyzed on the aspects of the contact angle range and its relationship with $G_w$. Different density ratios and surface tensions are applied to study their influences on the model.

### 3.1 Contact angle adjustment range with different density ratios

In previous studies, the density ratios are always small when simulating the contact angle. To study the influence of density ratio on the contact angle adjustment range, we simulated a stationary droplet on the solid surface for different density ratios and contact angles. The density ratios were chosen as 10, 100 and 1000. Parameters of the EOS were given in Table 1. The parameter $\kappa$ in Eq. (17) was chosen as 0.5, and parameter $A$ in Eq. (9) was given by -0.6. Initially, a semicircle droplet with radius of 30 was placed in the middle of the bottom wall. The relaxation matrix $\Lambda$ was given by $diag(0,1,1,1,1,1,1,1,1)$ in present work.

Simulation results show that when the density ratio is equal to 1000, the model is unstable for a necessary value of $G_w$ to attach the droplet to the bottom wall. Hence, here we only present the simulated contact angles when density ratios are equal to 10 and 100.

Table 1. Parameters of the EOS

| Density ratio | $\rho_1$ | $\rho_2$ | $\theta_v$ | $\theta_m$ | $\theta_l$ |
|---|---|---|---|---|---|
| 10 | 1.122 | 9.54 | 0.25 | -0.02 | 0.3 |
| 100 | 1.486 | 94.65 | 0.1633 | -0.02 | 0.33333 |
| 1000 | 1.325 | 971.1017 | 0.2 | -0.01 | 0.33333 |

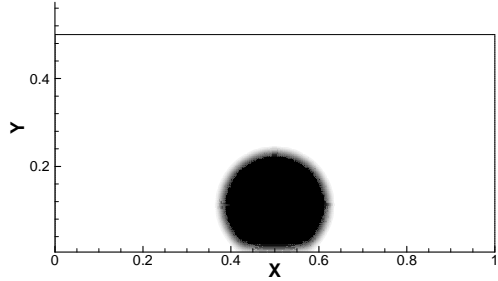

(a) $G_w = -0.766$

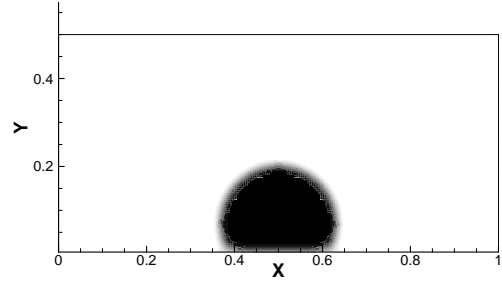

(b) $G_w = -1.1$

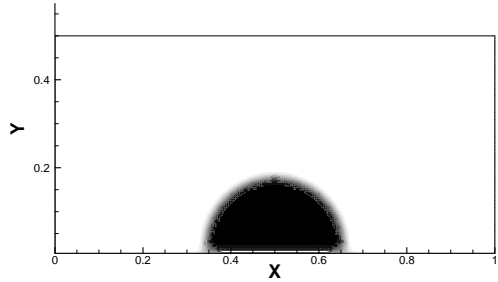

(c) $G_w = -1.43$

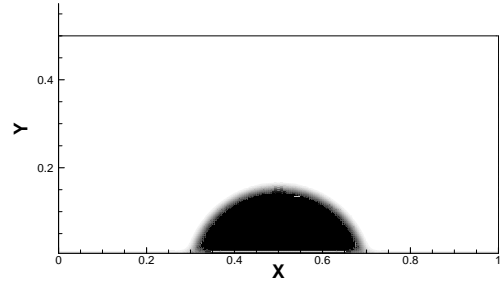

(d) $G_w = -1.77$

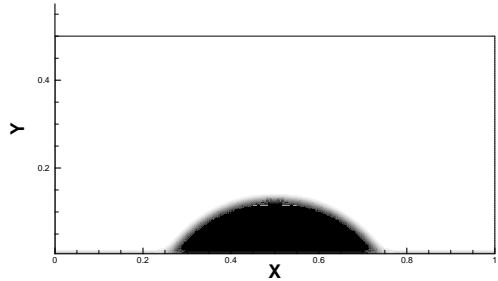

(e) $G_w = -2.1$

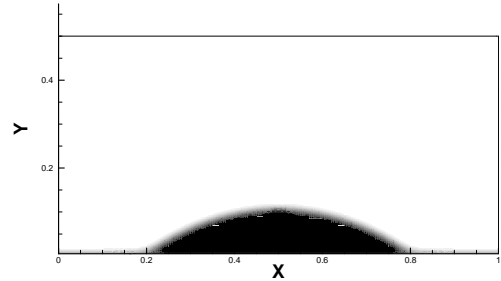

(f) $G_w = -2.43$

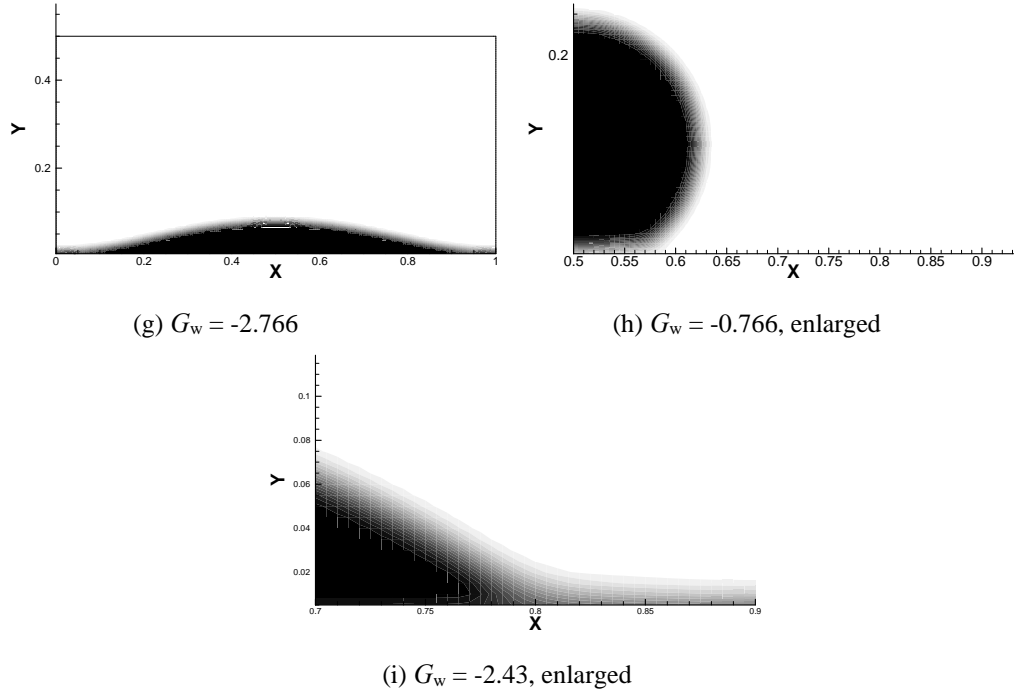

(g) $G_w = -2.766$    (h) $G_w = -0.766$, enlarged

(i) $G_w = -2.43$, enlarged

Fig. 1. Stationary droplet on the surface with diffident $G_w$ when density ratio is 10.

Figure 1 shows the simulation results of the stationary droplet on the surface with different values of $G_w$ when density ratio is 10. As we can see in these pictures, the contact angle can be adjusted nearly in a range of 0° to 180° by changing the value of $G_w$. However, it can be seen that the density of the droplet is not uniform near the wall when the contact angle is larger than 90° (Fig. 1 (h)), also the density of the vapor is condensed near the wall when the contact angles are small (Fig. 1 (i)). These results show that the contact angle can be adjusted in a large range when the density ratio is small, but the additional fluid-solid force influences the density distribution near wall.

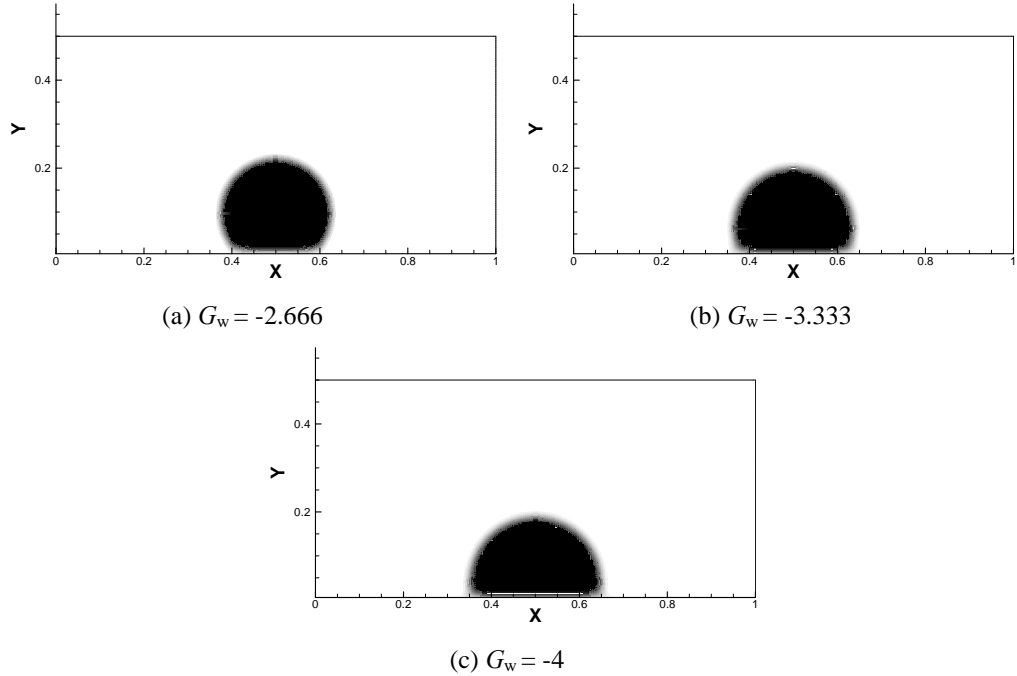

(a) $G_w = -2.666$     (b) $G_w = -3.333$

(c) $G_w = -4$

Fig. 2 stationary droplet on the surface with diffident $G_w$ when density ratio is 100.

The contact angel adjustment is also simulated for the density ratio set as 100. Fig. 2 shows the simulation results of the stationary droplet on the surface with different $G_w$ when density ratio is 100. The lowest $G_w$ we can get here is -4, and the corresponding contact angle is about 90° (Fig. 2 (c)). Contact angles smaller than 90° cannot be obtained due to the instability of the model. Moreover, the influence on the density near the wall is more significant, especially when the contact angle is large.

It can be seen from the above results that the density distribution near the wall is easily influenced by the solid-fluid interaction force. To clearly demonstrate this influence, the density distribution along the normal direction of the wall for density ratio equal to 10 is presented in Fig. 3. It can be seen from this figure that the densities change dramatically along the normal direction of the wall. Since the influence of the solid-fluid force is uncontrollable and difficult to evaluate, it may lead to many problems in practice.

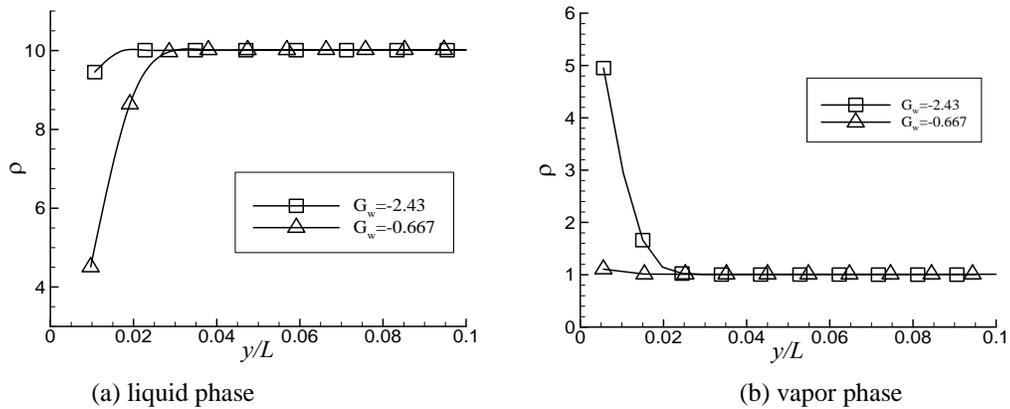

(a) liquid phase    (b) vapor phase

Fig. 3 The influence of Martys and Chen's contact angle adjustment method on the density distribution: (a) liquid phase (b) vapor phase.

In this section, we studied the contact angle adjustment with the solid-fluid interaction force method proposed by Martys and Chen for different density ratios. According to the results, we can find that the stability of the model is obviously influenced by the contact angle adjustment method especially when the density ratio is large. Moreover, a diffuse interface is artificially created between the bulk phase fluid and solid wall, which may influences the droplet movement on the wall and lead to more uncontrollable elements in the model.

**3.2 The influence of surface tension on the contact angle**

According to Young's equation (Eq. (18)), the contact angle is related to the surface tension of the fluids, if $\sigma_{sl}$, $\sigma_{sv}$ are constant, $\cos\theta_1$ should be inversely proportional to $\sigma_{lv}$. Since the surface tension $\sigma_{lv}$ is proportional to $1-\kappa$, the $\cos\theta_1$ should also be inversely proportional to $1-\kappa$. However, the mechanism of the contact angle adjustment method proposed by Martys and Chen is too complicate to theoretic analyze based on Young's equation. Hence, to study the influence of the surface tension on the contact angle, we simulated contact angles with different surface tensions, and the contact angle changing is invested based on the simulation results.

To maintain the stability of the model, the density ratio was given by 10, and the corresponding parameters were given in Table 1. Initially, a semicircle droplet with radius of 30 was placed in the middle of the bottom wall. The value of $G_w$ was given by $-1.43$, the corresponding contact angle was about $90°$ when $\kappa = 0.9$. To study the influence of surface tension on this model, we adjusted the surface tension by changing parameter $\kappa$. The simulation results are shown in Fig. 4.

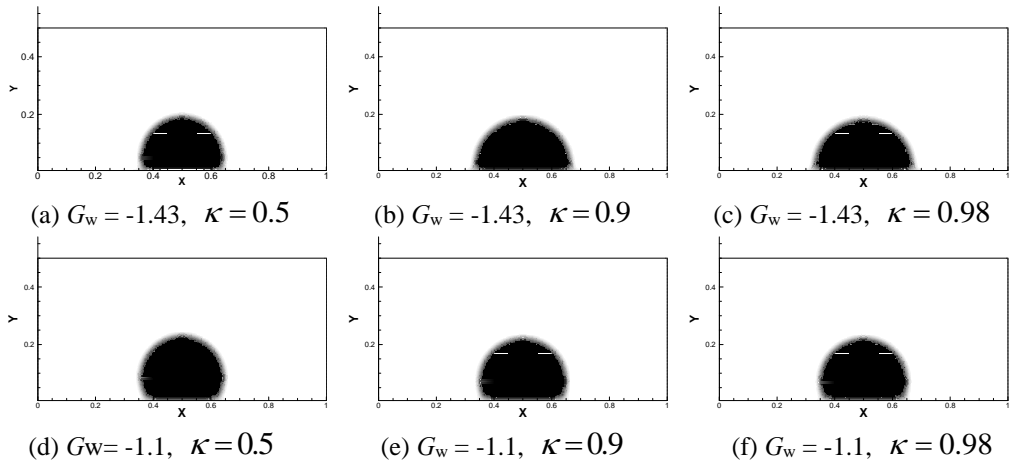

(a) $G_w = -1.43, \kappa = 0.5$     (b) $G_w = -1.43, \kappa = 0.9$     (c) $G_w = -1.43, \kappa = 0.98$

(d) $G_w = -1.1, \kappa = 0.5$      (e) $G_w = -1.1, \kappa = 0.9$      (f) $G_w = -1.1, \kappa = 0.98$

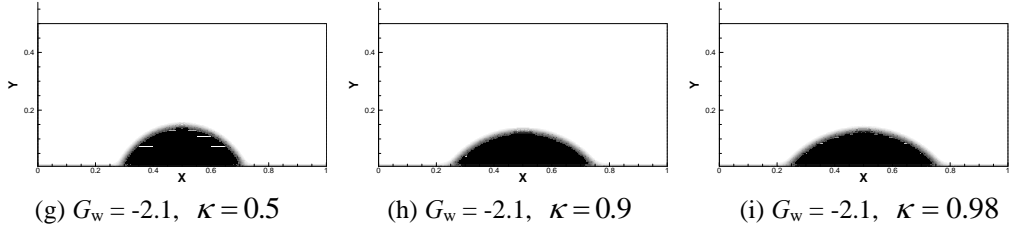

(g) $G_w = -2.1$, $\kappa = 0.5$    (h) $G_w = -2.1$, $\kappa = 0.9$    (i) $G_w = -2.1$, $\kappa = 0.98$

Fig. 4. Stationary droplet on the surface with diffident $\kappa$ when density ratio is 10.

As we can see in Fig. 4, the contact angle is slightly influenced by the value of parameter $\kappa$. However, the cosine function of the contact angle does not change proportionally with $1/(1-\kappa)$, which means that the surface tensions between the solid and liquids ($\sigma_{sl}$, $\sigma_{sv}$) are also influenced by $\kappa$. Hence, in practice, to simulate a certain contact angle, if the surface tension is changed, the parameter $G_w$ needs change too, which makes this method more inconvenient to be applied.

According to the analysis above, the contact angle adjustment method with solid-fluid interaction force has two main drawbacks: first, the method influences the stability of the pseudo-potential model, as a consequence, the contact angle adjustment range is small for large density ratio; in the second, the contact angle cannot be prescribed clearly, and the influence of surface tension on the contact angle is irregular, which makes this method more inconvenient to be applied.

**4 Geometrical contact angle adjustment method for pseudo-potential model.**

To overcome the drawbacks of Martys and Chen's contact angle adjustment method, here we introduced the geometrical contact angle adjustment method. As the name suggest, this method is based on the geometrical properties of the model. To adjust the contact angle, we add a layer of ghost cells adjacent to the solid boundary. The geometric formulation of the densities on the ghost lattices is given by [24]

$$\rho_{i,0} = \rho_{i,2} + \tan\left(\frac{\pi}{2} - \theta\right)\left|\rho_{i+1,1} - \rho_{i-1,1}\right|, \tag{19}$$

where the first and second subscripts denote the coordinates along and normal to the solid boundary, respectively. In the normal direction, 0 and 1 represent the ghost nodes and boundary nodes, respectively. However, the values of density directly influence the interaction interparticle force for pseudo-potential model, and the absolute value of $\tan\left(\frac{\pi}{2} - \theta\right)$ can be very large when $\theta$ get close to the zero or $\pi$. Consequently, the model may be unstable because of the rapidly changed interaction force near the solid boundary. To eliminate these effects of $\tan\left(\frac{\pi}{2} - \theta\right)$, we proposed a modified geometric formulation for the pseudo-potential model, which is given by

$$\rho_{i,0} = \rho_{i,2} + \sin\left(\frac{\pi}{2} - \theta\right)\left[(\rho_{i+1,1} - \rho_{i-1,1})^2 + (\rho_{i,2} - \rho_{i,0})^2\right]. \tag{20}$$

Note the above equation is an implicit expression of $\rho_{i,0}$. To simplify this equation, we can approximately replace $(\rho_{i,2} - \rho_{i,0})$ by $2(\rho_{i,2} - \rho_{i,1})$. However, it may reduce accuracy by using $\rho_{i,1}$ instead of $\rho_{i,0}$. To maintain the accuracy of the model, the iteration method is applied here to calculate $\rho_{i,0}$, which is given by

$$\rho_{i,0,1} = \rho_{i,2} + \sin\left(\frac{\pi}{2} - \theta\right)\left[(\rho_{i+1,1} - \rho_{i-1,1})^2 + 4(\rho_{i,2} - \rho_{i,1})^2\right]$$

$$\rho_{i,0,2} = \rho_{i,2} + \sin\left(\frac{\pi}{2} - \theta\right)\left[(\rho_{i+1,1} - \rho_{i-1,1})^2 + (\rho_{i,2} - \rho_{i,0,1})^2\right]$$

$$\rho_{i,0,3} = \rho_{i,2} + \sin\left(\frac{\pi}{2} - \theta\right)\left[(\rho_{i+1,1} - \rho_{i-1,1})^2 + (\rho_{i,2} - \rho_{i,0,2})^2\right]$$

$$\ldots \tag{21}$$

After the iteration, the solution of Eq. (20) can be approximately obtained.

**4.2 Numerical test of geometric contact angle adjustment method**

In this section, we numerically investigate the accuracy and stability of the new contact angle method. Since the stability of the model is easy to maintain when the density ratio is equal to 10, here we only consider the cases when density ratios are 100 and 1000. The parameters of the EOS are also given in Table 1, and the other parameters are the same as in section 3.1. After access the stability of the method, the influence of surface tension is also investigated and numerically compared with the original contact angle adjustment method.

**4.2.1 Contact angle adjustment range**

First we compare the results with and without the iteration. Figure 5 shows the results when the contact angle is set to 15°. Figure 5 (a) shows the simulation result with the density of ghost nets calculated only by Eq. (19). It can be seen from this figure that the simulated contact is about 30°, which is obviously larger than 15°. Figure 5 (b) shows the contact angel obtained after four times iteration. The contact angle of this figure is 15.2°, which is much closer to 15° compared with Fig 5 (a). Since the contact angle changes little for more iterations, we apply four times iteration in the following simulations.

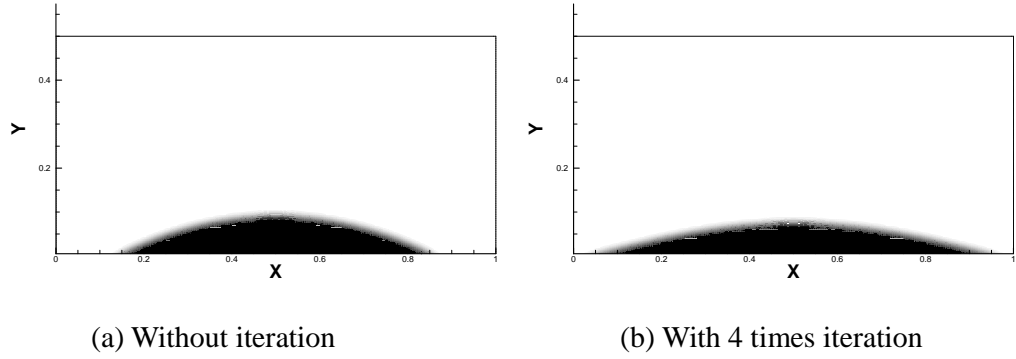

(a) Without iteration  (b) With 4 times iteration

Fig. 5. Contact angle obtained different iterations: (a) with 0 times iteration (b) with 4 times iteration

Figure 6 and Figure 7 show the contact angles when the density ratios are equal to 100 and 1000, respectively. It can be seen from these figures that a large range of contact angles (nearly 0° to 180°) can be obtained without influencing the stability of the model.

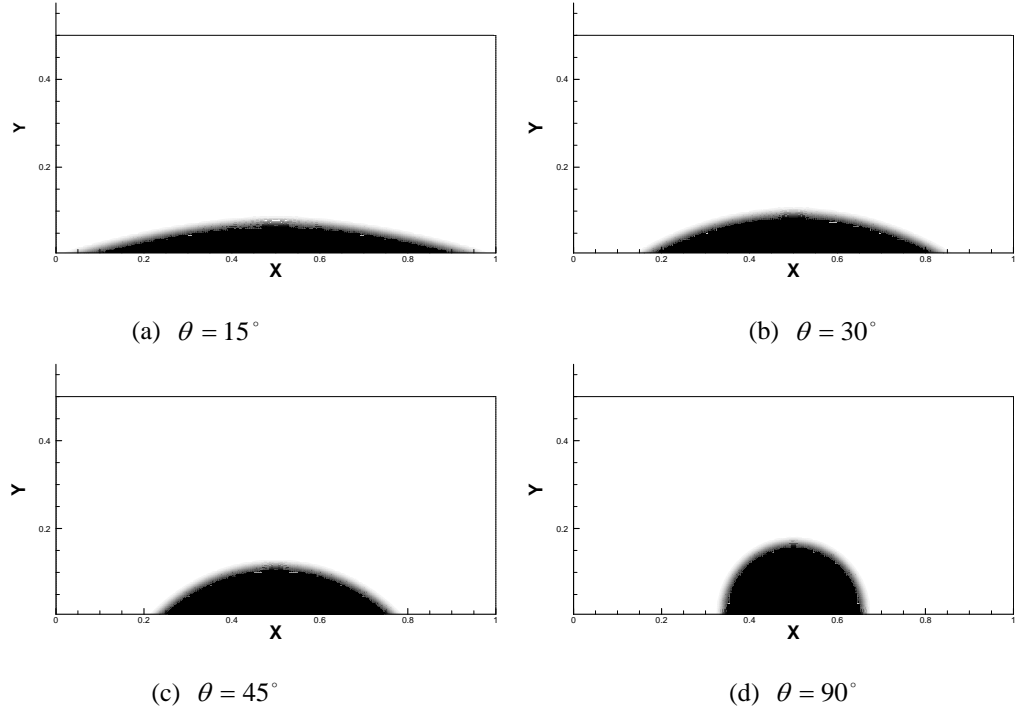

(a) $\theta = 15°$  (b) $\theta = 30°$

(c) $\theta = 45°$  (d) $\theta = 90°$

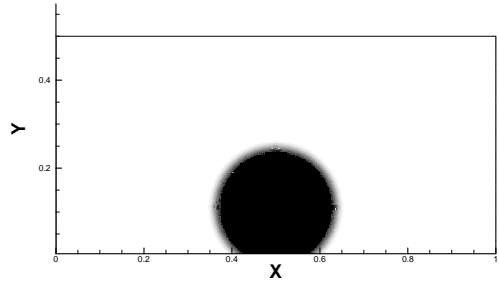

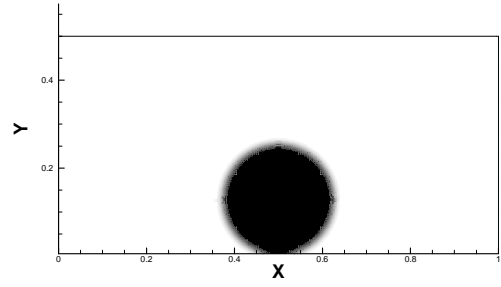

(e) $\theta = 135°$            (f) $\theta = 150°$

Fig. 6. Contact angles obtained with the geometry method when the density ratio is equal to 100.

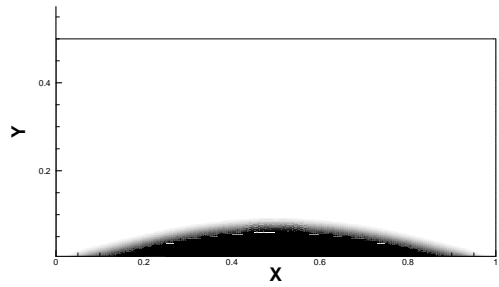

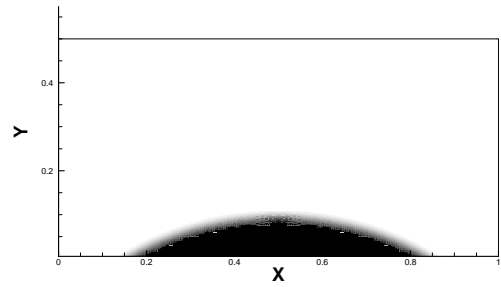

(a) $\theta = 15°$            (b) $\theta = 30°$

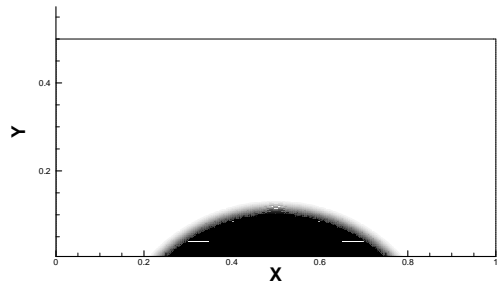

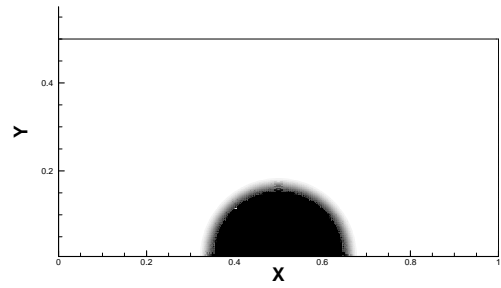

(c) $\theta = 45°$            (d) $\theta = 90°$

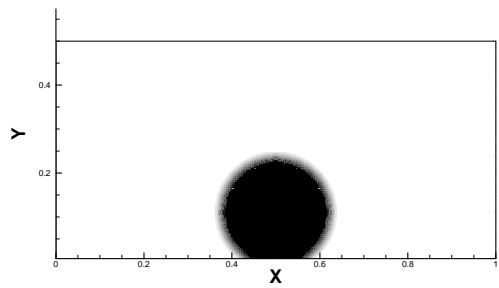

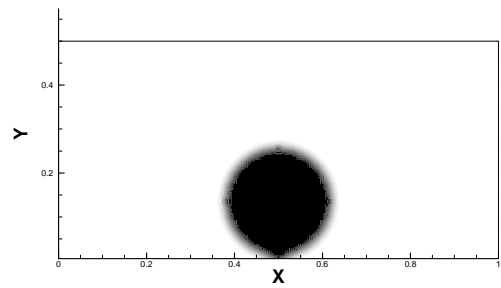

(e) $\theta = 135°$　　　　　　　　　　　　　(f) $\theta = 150°$

Fig. 7 Contact angles obtained with the geometry method when the density ratio is equal to 1000.

Figure 8 shows the density distribution along the normal direction of the wall for the new method when the contact angles are set to 45° and 135°. It can be seen from these figures that the density distributions are independent on the distance from the wall. Hence, the properties of the fluid are consistent with or without considering the influence of the wall. These results show that the new geometry method overcomes the drawbacks of Martys and Chen's model which reduces the stability of pseudo-potential model and artificially influences the density distribution of the fluid near the wall.

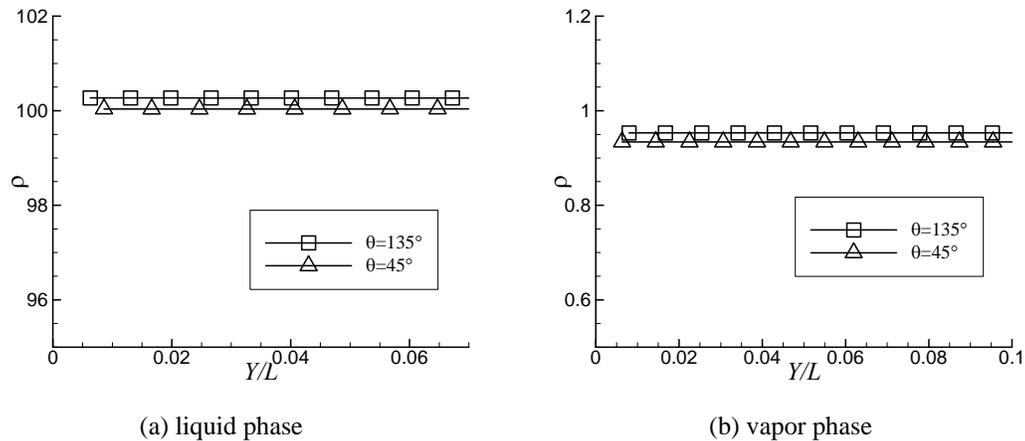

(a) liquid phase　　　　　　　　　　　(b) vapor phase

Fig. 8 The influence of geometry contact angle adjustment method on the density distribution: (a) liquid phase (b) vapor phase.

**4.2.2 Surface tension influence**

We further investigated the influence of the surface tension on the simulated contact angle. The simulation results of contact angles for different surface tensions when the density ratio is 100 are shown in Fig. 9. Measuring results show that the simulated contact angles are 46°, 90°, 134° for $\kappa = 0.9$, and 45°, 90°, 137° for $\kappa = 0.98$, which are very close to the set angles. Hence conclusion can be made that the proposed contact angle adjustment method is independent on the surface tension. However, the shape of the droplet slightly changes when $\kappa = 0.98$, a probable reason is that the surface tension in this case is too low to maintain the shape of the droplet under the influence of boundary interaction forces. These results show that the contact angle can be adjusted independently of the surface tension with the geometry method, hence, it could be more convenient to apply the pseudo-potential model with this contact angle adjustment method.

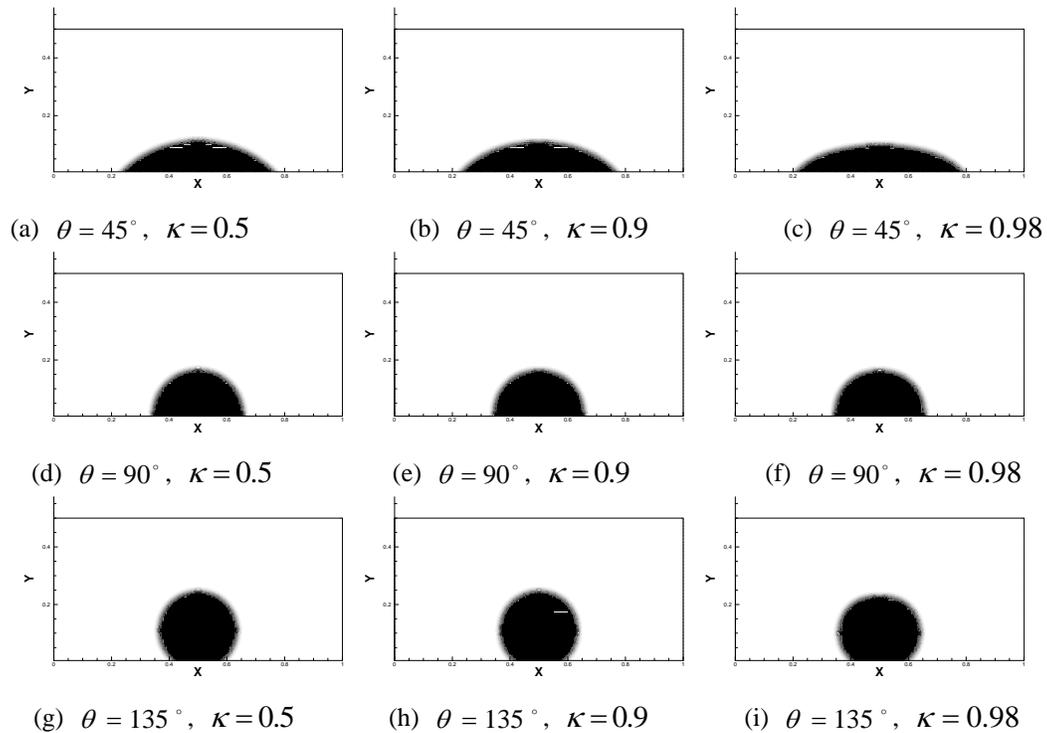

(a) $\theta = 45°$, $\kappa = 0.5$    (b) $\theta = 45°$, $\kappa = 0.9$    (c) $\theta = 45°$, $\kappa = 0.98$

(d) $\theta = 90°$, $\kappa = 0.5$    (e) $\theta = 90°$, $\kappa = 0.9$    (f) $\theta = 90°$, $\kappa = 0.98$

(g) $\theta = 135°$, $\kappa = 0.5$    (h) $\theta = 135°$, $\kappa = 0.9$    (i) $\theta = 135°$, $\kappa = 0.98$

Fig. 9. Contact angles obtained with the geometry method for different surface tensions

**5 conclusion**

In this paper, we numerically investigated the contact angle adjustment methods for the pseudo-potential model. Several drawbacks of the original contact angle adjustment method proposed by Martys and Chen were pointed out based on the simulation results. To overcome these drawbacks, a modified geometry contact angle adjustment method was proposed. Compared with the Martys and Chen's method, the presented method has the fellow advantages:

(a) The contact angle of the present method can be prescribed, and the accuracy of the model is also improved.

(b) The contact angle adjustment range is increased compared with the original method for moderate density ratio, and it can be applied when the density ratio is large without influencing the stability of the model.

(c) It overcomes the drawback of the original method which artificially changes the density distribution near the wall.

(d) The contact angle adjustment method is surface tension independent, hence, it is easier to be adjusted than the original method.

Overall, the presented geometry method is much improved compared with the original method proposed by Martys and Chen. It is expected that the present work will help setting the stage for future and more challenging applications of single component pseudo-potential methods in the simulation of complex multiphase.

**Acknowledgements**